\documentclass[11,runningheads,a4paper]{llncs}

\usepackage[sort&compress,numbers]{natbib}
\usepackage{comment}

\usepackage{amssymb,amsmath}
\usepackage{graphicx}
\usepackage{subcaption}

\usepackage{url}
\urldef{\mailsa}\path|beka@cmu.edu, {sventura, msadinle, fienberg}@stat.cmu.edu|

\usepackage{arabtex}
\usepackage{utf8}
\usepackage{soul}

\newcommand{\proba}[1]{\operatorname{P}\left( #1 \right)}
\newcommand{\npop}{N_{\text{pop}}}

\usepackage[switch]{lineno}
\RequirePackage[colorlinks,citecolor=blue,urlcolor=blue]{hyperref}
\usepackage[ruled,lined]{algorithm2e}
\SetKw{KwSet}{Set}

\begin{document}

\mainmatter  

\title{Generalized Bayesian Record Linkage and Regression with Exact Error Propagation}

\titlerunning{Generalized Bayesian Record Linkage and Regression}

\author{Rebecca C. Steorts$^{1}$, Andrea Tancredi$^{2}$,  and Brunero Liseo$^{2}$ %
 }
 
\institute{
$^{1}$Duke University and U.S. Census Bureau\\
Department of Statistical Science, affiliated faculty, Computer Science, Biostatistics and Bioinformatics, the information initiative at Duke (iiD), and the Social Science Research Institute (SSRI) Duke University; Principal  Researcher, Center for Statistical Research Methodology\\
$^{2}$La Sapienza, Rome, Italy\\
Department of Methods and Models for Economics, Territory and Finance \\
\path|{beka}@stat.duke.edu,{andrea.tancredi, brunero.liseo}@uniroma.it |
}

\authorrunning{Steorts, Tancredi, and Liseo}

\maketitle

\begin{abstract}
Record linkage (de-duplication or entity resolution) is the process of merging noisy databases to remove duplicate entities. While record linkage removes duplicate entities from such databases,  \emph{the downstream task} is  any inferential, predictive, or post-linkage task on the linked data. 
One goal of the downstream task is obtaining a larger reference data set, allowing one to perform more accurate statistical analyses. In addition, there is inherent record linkage uncertainty passed to the downstream task. Motivated by the above, 
we propose a generalized Bayesian record linkage method and consider multiple regression analysis as the downstream task. 
Records are linked via a random partition model, which allows for a wide class to be considered. In addition, we jointly model the record linkage and downstream task, which allows one to account for the record linkage uncertainty exactly. Moreover, one is able to generate a feedback propagation mechanism of the information from the proposed Bayesian record linkage model into the downstream task. This feedback effect is essential to eliminate potential biases that can jeopardize resulting downstream task. We apply our methodology to multiple linear regression, and illustrate empirically that the ``feedback effect" is able to improve the performance of record linkage.

\end{abstract}

\section{Introduction}
\label{sec:intro}
Record linkage (de-duplication or entity resolution) is the process of merging noisy databases to remove duplicate entities. While record linkage removes duplicate entities from such databases,  \emph{the downstream task} is any inferential, predictive or post-linkage task on the linked data. In this paper, we propose a joint model for the record linkage and the downstream task of linear regression. 
Our proposed model can link records over an arbitrary number of databases (lists or files). We assume there is duplication within each database, known as ``duplicate detection." Our record linkage model can be expressed as a random partition model, which leads to a large family of distributions. Next, 
we jointly model the record linkage task and the downstream task (linear regression), which allows for the exact propagation of the record linkage uncertainty into the downstream task. Crucially, this generates a feedback propagation mechanism from the proposed Bayesian record linkage model into the downstream task of linear regression. This feedback effect is essential to eliminate potential biases that can jeopardize resulting inference in the downstream task. We apply our methodology to multiple linear regression, and illustrate empirically that the ``feedback effect" is able to improve performance of record linkage.
\subsection{Prior Work}
Our work builds off \cite{tancredi:liseo2011, steorts14smered, steorts??bayesian, sadinle2014detecting}, which all proposed Bayesian record linkage models well suited for categorical data. 
\cite{tancredi:liseo2011} modeled the fully observed records through the  ``hit-and-miss" measurement error model \cite{copas:hilton}. One natural way 
to handle record linkage uncertainty is via a joint model of the record linkage and downstream task. \cite{liseo:tancredi2011} introduced a record linkage model for continuous data based on a multivariate normal model with measurement error.
Turning to just record linkage tasks, \cite{steorts14smered, steorts??bayesian} were the first to perform simultaneous record linkage and de-duplication on multiple files by using the fully observed records, creating a scalable record linkage algorithm. In similar work, de-duplication in a single database framework was tackled from a Bayesian perspective in \cite{sadinle2014detecting} by using the information provided by the comparison data. 

Related work regarding the record linkage and downstream task has been considered under specific  assumptions.  \cite{larsen:lahiri} assumed that the two databases represent a permutation of the same database of units and proposed an estimator (LL) of the regression coefficients which is unbiased, conditionally on the matching probabilities provided by the record linkage task. \cite{hof12} extended this approach to handle more complex and realistic linkage scenarios and logistic regression problems. Generalizations of the LL estimator have been also provided by \cite{kim12} using estimating equations. In addition, \cite{goldstein12} proposed  to consider the probabilities of being a match --- provided by the record linkage algorithm --- as an ingredient to be used within a multiple imputation scenario. Finally, \cite{gutman13} proposed a Bayesian method that jointly models the record linkage and the association between the overlapping features in two different databases. The authors consider somewhat simpler situation where the number of records to match in the two databases is relatively small and relies upon a specific blocking criteria. In addition, one potential limitation of the approach is the assumption of specific matching pattern.  For each single block of comparisons, all cases in the smaller database will certainly appear in the other databases. We refer to \cite{gutman2015error} for details.

Section \ref{sec:model} introduces our Bayesian record linkage model, providing extensions to priors on random partitions. Section \ref{sec:regression} generalizes our record linkage methodology to the downstream task of linear regression. 
Section \ref{sec:application} provides experiments for the record linkage task on synthetic data. We then provide three experiments on the joint record linkage and downstream task of linear regression on synthetic data. Section \ref{sec:discussion} provides a discussion and extensions to future work.
\section{Bayesian Record Linkage and Priors on Partitions}
\label{sec:model}
In this section, we introduce notation used through the paper, our Bayesian record linkage model, and an alternative and more intuitive construction for the prior on co-referent records, known as the linkage structure $\lambda.$
\subsection{Notation} 
Assume $L$ databases (lists, data sets, or files) $F_1, F_2\ldots, F_L$ that consist of either qualitative and/or categorical records, which are noisy due to the data collection process. Each record corresponds to an underlying latent entity (statistical unit) of partially overlapping samples (or populations). In addition, assume all databases have $p$ overlapping features (fields).
Assume 
$L$ sets of records are collected from a given population of size $\npop$ where $1\leq \npop \leq \infty$ in the same framework as \cite{steorts??bayesian, steorts2015entity}. As such, assign a label $j'$ ($j'=1,\ldots, \npop$) to each member of the population. Next, let $\tilde{v}_{j'}=(\tilde{v}_{j'1},\ldots, \tilde{v}_{j' p}$) be the vector of the $p$ categorical overlapping features for  the population individual $j'$. Finally, 
denote the entire set of population records by $\tilde{v}=(\tilde{v}_1,\ldots, \tilde{v}_{N_{pop}}).$ 
\subsection{Bayesian record linkage model}
\label{sec:model2}
Assume the set of population records $\tilde{v}$  is generated independently, for $j'=1,\ldots ,\npop$, from a vector of  independent categorical variables $\tilde{V}=(\tilde{V}_1,\ldots, \tilde{V}_\ell,\ldots, \tilde{V}_p)$  such that $\tilde{V_l}\in \{v_{\ell\, 1},\ldots, v_{\ell \, M_\ell } \}$  and 
\begin{equation}
P\left( \tilde{V}_\ell=v_{\ell\, s}\right)=\theta_{\ell\, v_{\ell\, s}} \quad s=1,\ldots, M_\ell,
\label{vtrue}
\end{equation}
where $M_\ell$ is the number of categorical values for the $\ell$th feature.  At the sample level, assume that one does not observe the ``true'' population  values due to measurement errors. Thus, the observed records, which is a database of size $N_i$, $i=1, \dots, L$, consists of distorted versions of subsets of the vectors $\tilde{v}_{j'}.$ 
%
Let $v_{ij}=(v_{ij1}, \ldots, v_{ijp})$ denote the observed values for the $j$-th record of the $i$-th database, where $i=1,\ldots, L$ and $j=1,\ldots, N_i$. Denote 
 the observed records (across the $L$ databases) by $v=(v_{11}, \ldots, v_{1 N_1}, \ldots, v_{L 1}, \ldots v_{L N_L}).$
%
Next, let the set of latent indicator variables $\lambda_{ij}\in\{ 1,\,\ldots, \npop\}$ denote the unknown co-reference (matching) pattern between the observed records $v$ and the population records $\tilde{v},$ where $\lambda_{ij}=j'$ indicates that the population record $j'$  generated the observed record $v_{ij}$.\footnote{The relation $\lambda_{i j_1}=\lambda_{i j_2}$, with $j_1\neq j_2$, implies that records $j_1$ and $j_2$ of the $i$-th database represents are co-referent to the same population record. This is an instance of duplicate-detection within the same database. When $\lambda_{i_1 j_1}=\lambda_{i_2 j_2}$, with $i_1\neq i_2$, one has the usual record linkage framework with the same individual appearing in two different databases.}
In general, let $\lambda=( \lambda_{11}, \dots, \lambda_{1N_1}, \ldots, \lambda_{L1}, \dots, \lambda_{L N_L})$ denote the linkage structure. 

Next, we formalize the distortion mechanism when the population records are observed in the $L$ databases using the  \textit{hit-and-miss model} \cite{copas:hilton}.  Let $V_{ij\ell}$ be the random variable  that generates observed record $v_{ijl}$.  Assume that  $V_{ijl}\in \{v_{l\, 1},\ldots, v_{l \, M_\ell } \}$, that is, $V_{ij\ell}$ has the same support of $\tilde{V_\ell}$. Let $\delta_{a,b}=1$ if $a=b$ and $\delta_{a,b}=0$ if $a\neq b$, 
which implies that
\begin{equation}
\label{hitmiss}
P(V_{ij\ell}=v_{\ell \, s} \mid \lambda_{ij}, \tilde{v} ,\alpha_\ell )
=(1-\alpha_\ell) \delta_{\tilde{v}_{\lambda_{ij} \ell} ,v_{\ell\,s }}+\alpha_\ell \theta_{\ell \,v_{\ell\,s}} \quad s=1,\ldots, M_\ell
\end{equation}
for $i=1,\ldots, L; \, j=1,\ldots, N_i; \, \ell=1,\ldots, p,$
where
$\alpha_\ell \in [0,1]$ represents the distortion probability for the $\ell$-th overlapping feature.
Here, the true population value is observed with probability
$1-\alpha_\ell$, and  a different value is drawn from the random variable $\tilde{V}_\ell$ generating the population values with probability $\alpha_\ell.$
Finally, assuming the conditional independence among all the overlapping features given their respective unobserved population counterparts, one obtains
\begin{equation}
\label{model1}
p(v\mid \tilde{v}, \lambda, \alpha)=\prod_{i=1}^L \prod_{j=1}^{N_i} \prod_{\ell=1}^p
P(v_{ijl} \mid  \tilde{v},\lambda,\alpha) =\prod_{i=1}^L \prod_{j=1}^{N_i} \prod_{\ell=1}^p[ (1-\alpha_\ell) \delta_{\tilde{v}_{\lambda_{ij} \ell} ,v_{ij\ell}}+\alpha_\ell \theta_{\ell\, v_{ij\ell}}].
\end{equation}
We assume that the distortion probabilities are exchangeable, that is
$$\alpha_{\ell} \stackrel{iid}{\sim} \text{Beta}(f,g), \, \ell=1,\dots ,p,$$ and
we assume the probabilities $\theta_{\ell \,1} \ldots \theta_{\ell \, M_\ell}$ are considered known and equal to the corresponding population frequencies. 
%
%
The model summarized by equations (\ref{vtrue}) and (\ref{model1}) can be viewed as a latent variable model where the unobserved  population records $\tilde{v}$ generate the observed records $v$   and $\alpha=(\alpha_1\ldots,\alpha_p)$ can be viewed as the unknown model distortion  parameter. 

\textbf{Remark}: A convenient property of the hit-miss model is that one can integrate out the unknown population values $\tilde{v}$ to directly obtain the distribution $p(v|\alpha,\lambda)$. The resulting marginal distribution $p(v|\alpha,\lambda)$ is the product of within-cluster distributions. To improve mixing, we use a Metropolis within Gibbs algorithm to simulate from the joint posterior $p(\lambda,\alpha\vert v).$
(See Appendix \ref{app:metropolis}).

\subsection{The prior distribution for $\lambda$}
\label{prior}
In this section, we propose a more intuitive and subjective construction of a prior distribution on $\lambda.$
Let  $z$ denote the random partition of the observed records determined by $\lambda$ and let $\mathcal{P}$ denote the set containing all the possible partitions of the $N$ observed records. The distribution on the sample labels $\lambda$ induces a distribution on $\mathcal{P}$. Furthermore, matches and duplicates are completely specified given the knowledge of the random partition $z$, which is invariant with respect to the labelings of the partition blocks. Given this construction, one can directly focus on the partition distribution of the observed records without linking the labels distribution to a sample design and to a population size $\npop$, see for example, \cite{sadinle2014detecting}. One can effectively consider the distribution of $\lambda$ as a prior distribution for the latent linkage structure and concentrate only on its probabilistic properties. Both the interpretations of the role of $\lambda$ (either as a consequence of the sampling design or a model represented by partitions) may provide useful insights for a correct choice of its prior distribution.
%
%
%
One difficult and related question in the record linkage literature has been the subjective specification on the space of partitions. 
%
%
%
%
A simple, alternative  prior for the number of distinct entities $k(z)$ can be obtained  looking  at the following allocation rule for the record labels which is based on a generalization of the Chinese Restaurant Process, namely the Pitman-Yor process (PYP) (\cite{pitman}, \cite{deblasi}). (See Appendix \ref{app:pyp} for details).

\section{The downstream task of linear regression}
\label{sec:regression}
In this section, we propose record linkage methodology for the downstream task of linear regression. 
Consider the model $\tilde{Y}=\sum_{l=1}^p \tilde{X}_{l}\beta_l+\epsilon$ for the population units, where the goal is to estimate the regression coefficients $\beta=(\beta_1,\ldots, \beta_p)^t$.   
%
We observe
$Y$ and $X=(X_1,\ldots, X_p),$
where $X$  represents a noisy measurement of the true covariates $\tilde{X}=(\tilde{X}_1,\ldots \tilde{X}_p)$ and $Y$ is a random copy of the corresponding population variable $\tilde{Y}$. 

To better illustrate our approach, we consider two scenarios. In the first scenario---\emph{the complete regression scenario}---each database reports a set of overlapping features, the response variable, and the covariates.
Let $y_{ij}$ and $x_{ij}=(x_{ij1}\ldots,x_{ijp})$ denote the observed
values for the $j$-th unit of the $i$-th database, where $i=1,\ldots, L$
and $j=1,\ldots, N_i$. In addition, let  $(y,x)$ denote the entire set of
regression data observed across the $L$ databases. In the \emph{complete scenario},
there is not a bias problem concerning the estimation of the $\beta$ coefficients. 
In the second scenario---\emph{the broken regression scenario}---we assume that the overlapping features are observed in each database, the response variable is observed in only the first database, and specific subsets of covariates are observed in the other
databases. In this situation, let $(y,x)$ denote the observation $y_{1j}$, where
$j=1,\ldots,N_1$ and $x_{ij}$, where $i=2,\ldots, L$ and $j=1,\ldots,
N_i$. Note that $x_{ij}$ represents only a fixed subset of the values
$x_{ij1}\ldots x_{ijp}$ for $j=1,\ldots,N_i$. Here, there is a bias issue regarding estimating the $\beta$ coefficients.\footnote{In both scenarios, we assume that the covariates have zero mean
and the regression model does not have an intercept.}


\subsection{Simple linear regression}
\label{sec:regression-simp}
In this section, we consider linear regression and the two scenarios mentioned above with a single covariate $X.$ First, consider the \emph{complete regression scenario}. 
Let $\tilde{X}_{j'}$ be  the true value of observation $X$ corresponding to the records of cluster  $C_{j'}$. Now consider a cluster $C_{j'}=\{(i,j)\}$ with one record.  Given the true value of $\tilde{X}_{j'}=\tilde{x}_{j'}$ and membership to cluster $C_{j'}$, we assume that the response variable $Y_{ij}$ follows a standard normal regression model with covariate $\tilde{x}_{j'}$, where the observed value for the covariate $X_{ij}$ is normal with mean $\tilde{x}_{j'}$ and $Y_{ij}$ and $X_{ij}$ are independent. That is,
\begin{equation}
\left[
\begin{array}{c}
Y_{ij}\\
X_{ij}
\end{array}
\right] \left| \right. \tilde{X}_{j'}=\tilde{x}_{j'}\sim N_2 \left[
\left(
\begin{array}{c c}
\beta & 0\\
0 & 1
\end{array}
\right)
\left[ \begin{array}{c} \tilde{x}_{j'}\\
\tilde{x}_{j'}\\
\end{array}
\right],
\left(
\begin{array}{c c}
\sigma^2_{y|\tilde{x}} & 0\\
0 & \sigma^2_{x|\tilde{x}}
\end{array}
\right)
\right] .
\label{yx1}
\end{equation}
We assume that
$\tilde{X}_{j'} \sim N(0, \sigma^2_{\tilde{x}})$, which
allows one to integrate $X_{j'}$ via equation~\ref{yx1}.
In fact, setting  $Z_{ij}= \left( Y_{ij},X_{ij} \right)^\prime$, one can easily show that 
conditionally
on the event $\{ (i,j)\in C_{j'}\}$, it follows that
\begin{equation}
Z_{ij} \sim N_2
\left[
\left(
\begin{array}{c}
 0\\
0
\end{array}
\right),
\sigma^2_{\tilde{x}}
\left(
\begin{array}{c c}
\beta^2 & \beta \\
\beta & 1 \\
\end{array}
\right)+
\left(
\begin{array}{c c}
\sigma^2_{y|\tilde{x}} & 0\\
0 & \sigma^2_{x|\tilde{x}}
\end{array}
\right)
\right].
\label{yx2}
\end{equation}
For ease of notation, let $I_{n}$ denote the $n\times n$ identity matrix, $0_n$ denote the $n$-vector of zero;
$1_{n}$ denote a vector of all $1$'s, and $J_n = 1_n 1_n^\prime$. Next, set
$$
B = \left(\begin{array}{c c}
\beta^2 & \beta \\
\beta & 1
\end{array} \right)
\quad
\mbox{and }
\quad
\Sigma = \left(\begin{array}{c c}
\sigma^2_{y \vert \tilde{x}}  & 0 \\
0 & \sigma^2_{x \vert \tilde{x}}
\end{array} \right).
$$
Consider a cluster $C_{j'}=\{ (i_1,j_1), (i_2,j_2) \}$ with two records.
The two pairs $Z_{i_1 j_1}$ and $Z_{i_2 j_2}$ are random vectors, both depending on the same ``true'' value $\tilde{X}_{j'}$. 
Let  $\otimes$ be the Kronecker product. 
Conditionally on $\tilde{X}_{j'}=\tilde{x}_{j'}$ and on the cluster membership, we replicate the model for a cluster with one record by assuming that  $Z_{i_1 j_1}$ and $Z_{i_2 j_2}$ are two independent bivariate normal random variables
with joint distribution 
\begin{equation} N_4 \left[
\left(I_2 \otimes
\left(
\begin{array}{c c}
\beta & 0\\
0 & 1
\end{array}
\right) \right)
\left(
1_4
\tilde{x}_{j'} \right) , I_2 \otimes \Sigma \right ].
\label{yx3}
\end{equation}
Then the marginal distribution of
$(Z_{i_1 j_1}, Z_{i_2 j_2} )^\prime$ is
$$\begin{pmatrix}
Z_{i_1 j_1} \\
Z_{i_2 j_2} \\
\end{pmatrix}
 \sim N_4 \big ( 0_4 , I_2 \otimes \Sigma + \sigma_{\tilde{x}}^2
 J_2 \otimes B \big ).
$$
This argument can be extended to any cluster size. When card$(C_{j^\prime}) = n$, the marginal distribution of $Z=(Z_{i_1 j_1}, \dots , Z_{i_n j_n })$ is again multivariate normal:
$
Z \sim N_{2n} \left ( 0_{2n}, I_n \otimes \Sigma + \sigma_{\tilde{x}}^2
J_n \otimes B \right ).
$

Next, consider the \emph{broken regression scenario}. In this case, when  some information is missing---either the covariate in the first database  or the response variable in some of the other databases---one can easily marginalize over the missing variables by using standard properties of multivariate normal distribution. 
Let  $(y,x)_{C_j'}=((y_{ij}, x_{ij} ) : \lambda_{ij}=j')$ denote the set of regression observations, which conditionally on $\lambda$, correspond to the $j^\prime$-th population unit.  For example, for a cluster $C_{j'}=\{(1,j)\}$ with one record in the first database,  we denote this as $(y,x)_{C_j'}=y_{1j}$. 
Using the marginal density of $Y_{ij}$  in equation~\ref{yx2}, we can write 
the likelihood, conditional on $\lambda$,  as $p((y,x)_{C_j'}|\lambda, \beta, \sigma^2_{y|\tilde{x}}, \sigma^2_{x|\tilde{x}}).$
Similarily, suppose $C_{j'}=\{(i,j)\}$ with $i>1$, then $(y,x)_{C_j'}=x_{ij}$ and the likelihood is given by marginal  density of $X_{ij}$. Next, consider a cluster $C_{j'}=\{ (1, j_1), (i_2,j_2) \}$ with a record in the first database and the other record in a different database, i.e. $i_2>1.$ It follows that 
 $(y,x)_{C_j'}=(y_{1j_1},x_{i_2 j_2})$ and the corresponding likelihood is found by marginalizing over the missing values 
$X_{1 j_1},Y_{2 j_2}$ in equation~\ref{yx3}, where we obtain the  joint density in equation~\ref{yx2}.
Finally,  it follows that the likelihood function (as a function of  $\lambda, \beta, \sigma^2_{y|\tilde{x}}, \sigma^2_{x|\tilde{x}}$) for both the complete and broken regression scenarios can be generally written as
$p(y,x| \lambda, \beta, \sigma^2_{x|\tilde{x}}, \sigma^2_{y|\tilde{x}})=\prod_{j'=1}^{N_{pop}} p( (y,x)_{C_j'}|\beta, \sigma^2_{x|\tilde{x}}, \sigma^2_{y|\tilde{x}}).$\footnote{
We assume that population units $j'$ that do not have an observed cluster size contribute to the likelihood with a factor equal to 1.}

In order to handle the record linkage and downstream regression task simultaneously, we assume conditional independence on $\lambda$ between the overlapping features in the record linkage model and the set of variables in the downstream task of linear regression. 
Assuming conditional independence, we find
\begin{eqnarray}\nonumber p(\lambda,\beta,\alpha, \sigma^2_{y|\tilde{x}}, \sigma^2_{x|\tilde{x}} |v,x,y)&\propto& p(v|\lambda,\alpha) p(y,x|\lambda,\beta,\sigma^2_{y|\tilde{x}}, \sigma^2_{x|\tilde{x}}) \\ &\times & p(\lambda) p(\alpha) p(\beta, \sigma^2_{y|\tilde{x}}, \sigma^2_{x|\tilde{x}}). 
\label{postreg}
\end{eqnarray}
The first factor is related to the record linkage process, and second factor is related to the downstream task of linear regression, and the other factors represent the prior distributions. We assume independent diffuse priors for $\beta, \sigma^2_{y|\tilde{x}}, \sigma^2_{x|\tilde{x}}$. To update the appropriate regression parameters $\beta, \sigma^2_{y|\tilde{x}}, \sigma^2_{x|\tilde{x}}$, we use the Metropolis-Hastings algorithm in Appendix \ref{app:metropolis}. Using the factorization of the posterior in equation~(\ref{postreg}), the proposed method can be generalized to any statistical model. 


\subsection{Multiple linear regression}
\label{sec:regression-m}
We extend the downstream task to that of  multiple regression, first considering the complete regression scenario. 
Let $C_{j'}$ denote a cluster of size $n$, $Y_{C_{j'}}$ denote a vector with $n$ observations of the response variable in this cluster,  and $X_{C_{j'}}$ denote the $n \times p$ matrix with the values of the $p$ covariates observed in the cluster units. Let $[YX]_{C_{j'}}$ denote the vector of $n (p+1)$ elements with the $n$ rows of the matrix $(Y_{C_ {j'}}, X_{C_{j'}})$ vertically stacked 
and let $\tilde{X}_{j'}$ denote the vector containing the true values of the $p$ covariates. Equation~\ref{yx1} can be generalized assuming that
 $$
[YX]_{C_{j'}} \left| \right. \tilde{X}_{j'}\,\sim N_{n(p+1)} \left[
\left(
 I_{n\times n} \otimes
\left(
\begin{array}{c c}
\beta^t & 0_{p}^t\\
0_{p\times p} & I_{p\times p}
\end{array}
\right)
\right) \left(
1_{2n} \otimes \tilde{X} \right),
I_{n\times n} \otimes
\left(
\begin{array}{c c}
\sigma^2_{y|\tilde{x}} & 0\\
0 & \Sigma_{x|\tilde{x}}
\end{array}
\right)
\right],
$$
where
$$ 1_{2n} \otimes \tilde{X}\sim N_{2 n p} \left(0_{2np}, (1_n 1_n^t )\otimes
\left(
\begin{array}{c c}
\Sigma_{\tilde{x}} & \Sigma_{\tilde{x}}\\
\Sigma_{\tilde{x}} & \Sigma_{\tilde{x}}
\end{array}
\right)
 \right).$$
This way the marginal distribution of $[YX]_{C_{j'}}$ is $n(p+1)$-variate normal with zero mean and covariance matrix
\begin{equation*}
\begin{gathered}
\left(
I_{n\times n} \otimes
\left(
\begin{array}{c c}
\beta^t & 0_{p}^t\\
0_{p\times p} & I_{p\times p}
\end{array}
\right) \right)
\left((1_n 1_n^t )\otimes
\left(
\begin{array}{c c}
\Sigma_{\tilde{x}} & \Sigma_{\tilde{x}}\\
\Sigma_{\tilde{x}} & \Sigma_{\tilde{x}}
\end{array}
\right)
\right)
\left(
I_{n \times n}\otimes
\left(
\begin{array}{c c}
\beta^t & 0_{p}^t\\
0_{p\times p} & I_{p\times p}
\end{array}
\right) \right)^t
+\\
\left(I_{n\times n} \otimes
\left(
\begin{array}{c c}
\sigma^2_{y|\tilde{x}} & 0\\
0 & \Sigma_{x|\tilde{x}}
\end{array}
\right)\right),
\end{gathered}
\end{equation*}
which simplifies into
$$
(1_n 1_n^t )\otimes
\left(
\begin{array}{c c}
\beta^t \Sigma_{\tilde{x}} \beta & \beta^t \Sigma_{\tilde{x}} \\
\Sigma_{\tilde{x}} \beta^t & \Sigma_{\tilde{x}}
\end{array}
\right) +
I_{n\times n} \otimes
\left(
\begin{array}{c c}
\sigma^2_{y|\tilde{x}} & 0\\
0 & \Sigma_{x|\tilde{x}}
\end{array}
\right).
$$
The likelihood provided by the multiple regression model is the product of the factors
$p([YX]_{C_{j'}}=[y,x]_{C_j'}|\beta, \sigma^2{y|\tilde{x}}, \Sigma_{x|\tilde{x}})$ for the observed clusters. The same considerations from linear regression regarding modeling the prior and the computational aspects apply to multiple linear regression. 
Note the major difference is in the marginalization pattern in the broken regression scenario. In fact, for a cluster joining records across more than one database, we may need to integrate out the covariate values missing in the databases that share a cluster.

%

\section{Experiments}
\label{sec:application}

To investigate the performance of our proposed methodology we consider the \texttt{RLdata500} data set from the  \texttt{RecordLinkage} package in \texttt{R}. This synthetic data set consists of 500 records, each comprising first and last name and full date of birth. We modify this data set to consider two databases, where each database contains 250 records, respectively, with duplicates in and across the two databases. To consider the case without duplicate detection, we modify the original \texttt{RLdata500} such that it has no duplicate records within each of the two databases. Without duplicate detection is a special case of our general methodology (see Appendix \ref{app:duplicate-detection}).  We provide experiments for both record linkage and the downstream task.

\subsection{Record linkage with and without duplicate-detection}
\label{sec:linkage-dd}
\begin{figure}[h]
\centerline{\includegraphics[width=0.6\textwidth]{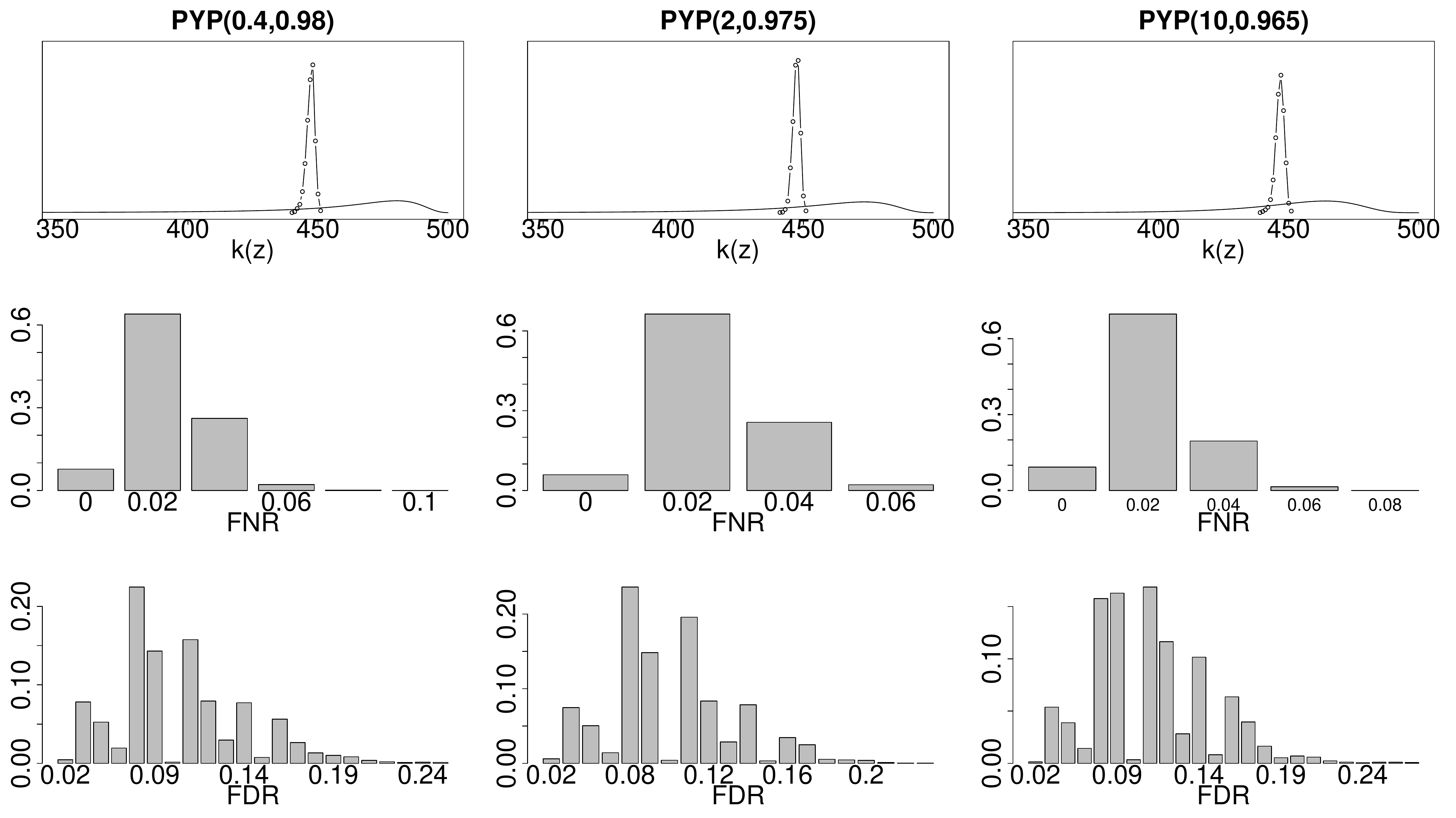}}
\caption{\label{fig2sis}
Prior and posteriors for $k(z)$ (first row), FNR posteriors (second row), FDR posteriors (third row) for the \texttt{RLdata500} data set. 
}
\end{figure}

\begin{figure}
\centerline{\includegraphics[width=0.6\textwidth]{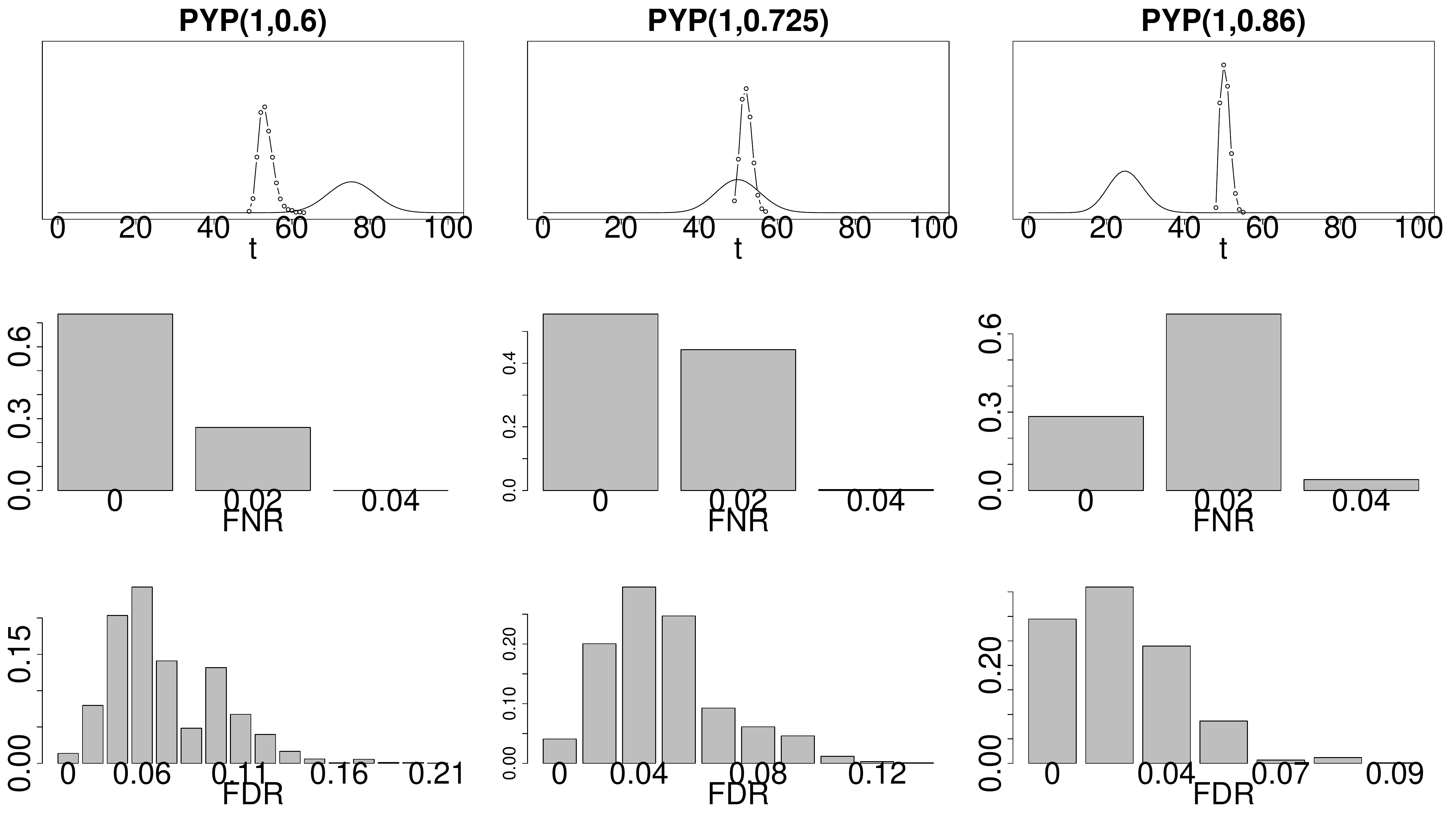}}
\caption{\label{fig4}
Prior and posteriors for $t$ (first row), FNR posteriors (second row), FDR posteriors (third row) for the \texttt{RLdata500} data set  assuming a two database record linkage framework without duplicate-detection. 
}
\end{figure}

We provide two record linkage experiments --- one with duplicate detection and one without duplicate detection. In Figures \ref{fig2sis} and Figure \ref{fig4}, we report the prior and the posterior for $k(z)$ and the performance of the record linkage procedure measured in terms of the posteriors of the false negative rates (FNR) and the false discovery rates (FDR). (For a review of FNR and FDR, see \cite{steorts2015entity, christen12data}). 

%
%
Figure \ref{fig2sis} (with duplicate detection)  illustrates that the resulting posteriors of $k(z)$ appears robust to the choices of $\theta$ and $\sigma$ (first row). We observe similar behavior for the posteriors of FNR and FDR (second and third rows). Figure \ref{fig4} illustrates that as we vary the PYP parameters,  the posterior of $T$ is weakly dependent on their values. The two database framework without duplicate detection leads, a posteriori,  to similar  FNR (second row)  and lower FDR (third row) compared to the previous case. (See Appendix~\ref{app:record-linkage} for the PYP parameter settings).



%
%

\subsection{Regression experiments}
\label{sec:experiments-regression}
We consider three regression experiments on the \texttt{RLdata500} data set. In Experiment I, we consider the complete regression scenario in a single database framework with duplicate detection. In Experiment II, we consider the broken  regression scenario with record linkage and duplicate detection. In Experiment III, we consider the broken multiple regression scenario in a  two database framework without duplicate detection. (See Appendix~\ref{app:regression} for details).

\begin{figure}[h]
\centerline{\includegraphics[width=0.85\textwidth]{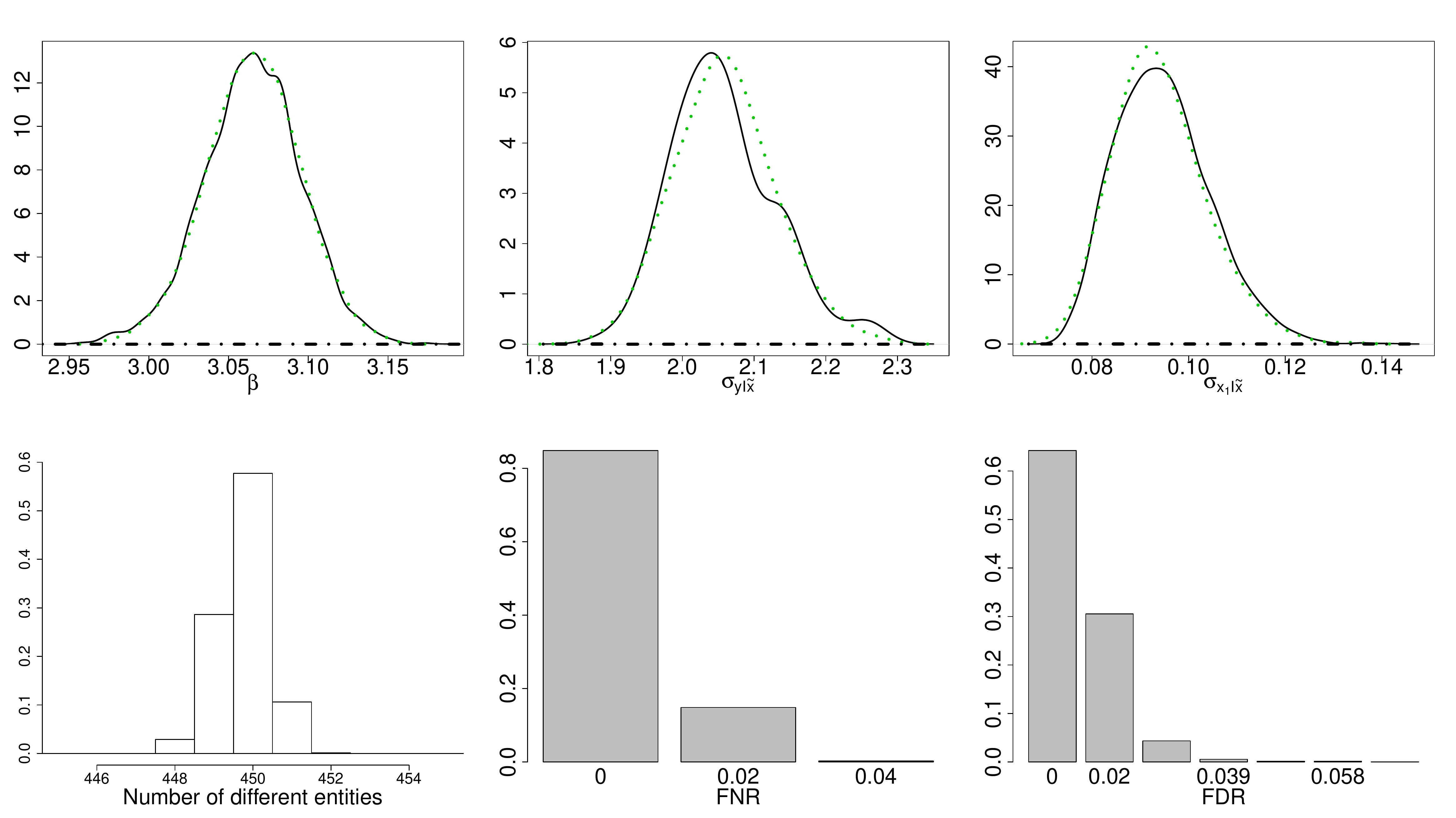}}
\caption{ \label{complete} 
Experiment I. Upper panels: prior (dotdash lines) and posterior  of $\beta, \sigma_{y|\tilde{x}}, \sigma_{x|\tilde{x}} $ with the joint record linkage and regression model (solid lines) and the true linkage structure (dotted lines).
Lower panels: posterior for $k(z)$,  FNR and FDR.}
\end{figure}

\begin{figure}[h]
\centerline{\includegraphics[width=0.85\textwidth]{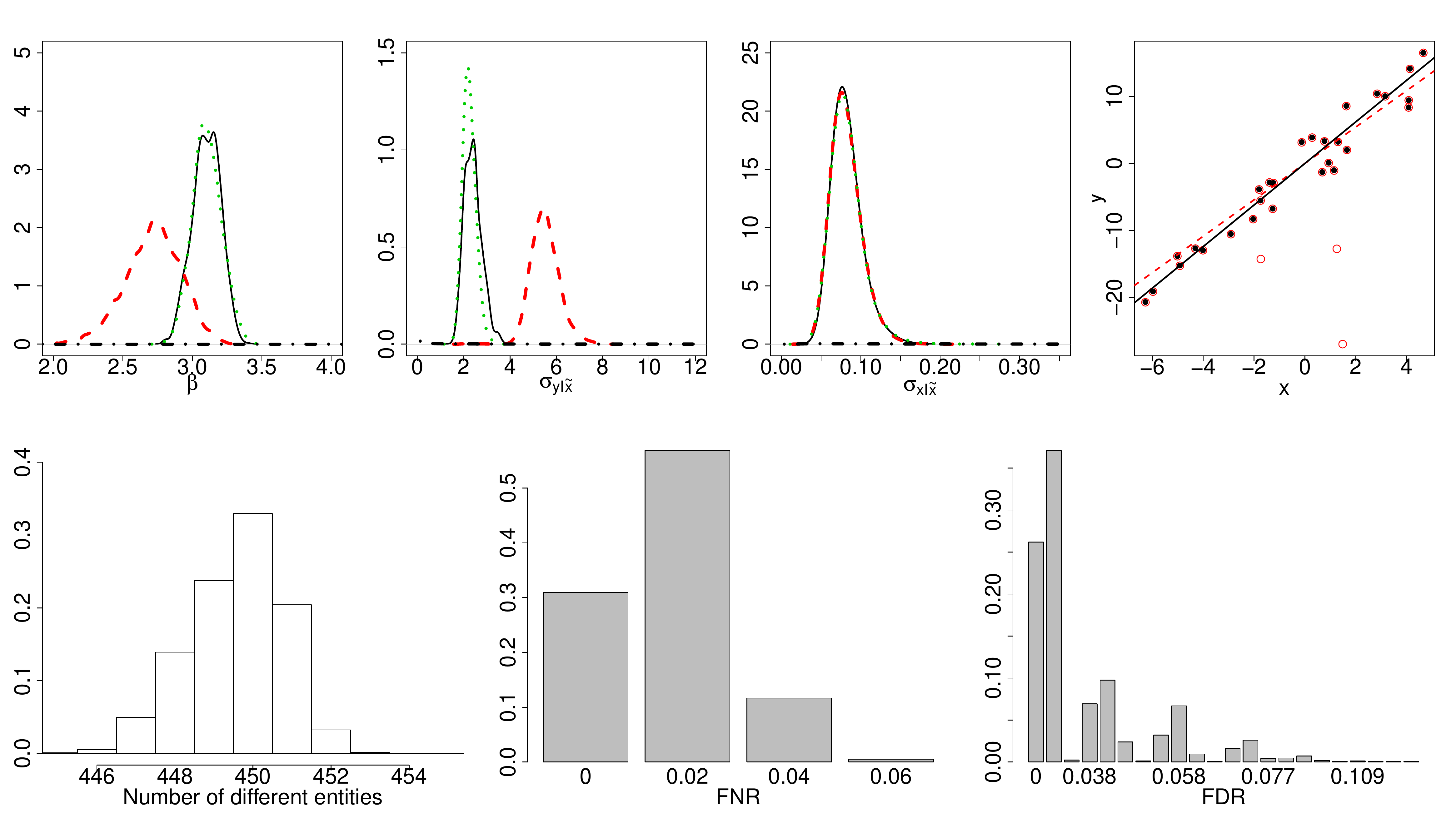}}
\caption{\label{broken}
Experiment II. Left upper panels: prior (dotdash lines) and posterior of $\beta,  \sigma_{y|\tilde{x}}, \sigma_{x|\tilde{x}} $ with the joint record linkage and regression model (solid lines), the true linkage structure (dotted lines) and the plug-in approach (dashed lines). Right upper  panel: estimated regression line and $(x,y)$ pairs with the  joint model (solid line and full circles) and the  plug-in approach (dashed line and empty circles) 
Lower panels: posterior  for $k(z)$, FNR and FDR.}
\end{figure}

\begin{figure}[h]
\centerline{\includegraphics[width=0.85\textwidth]{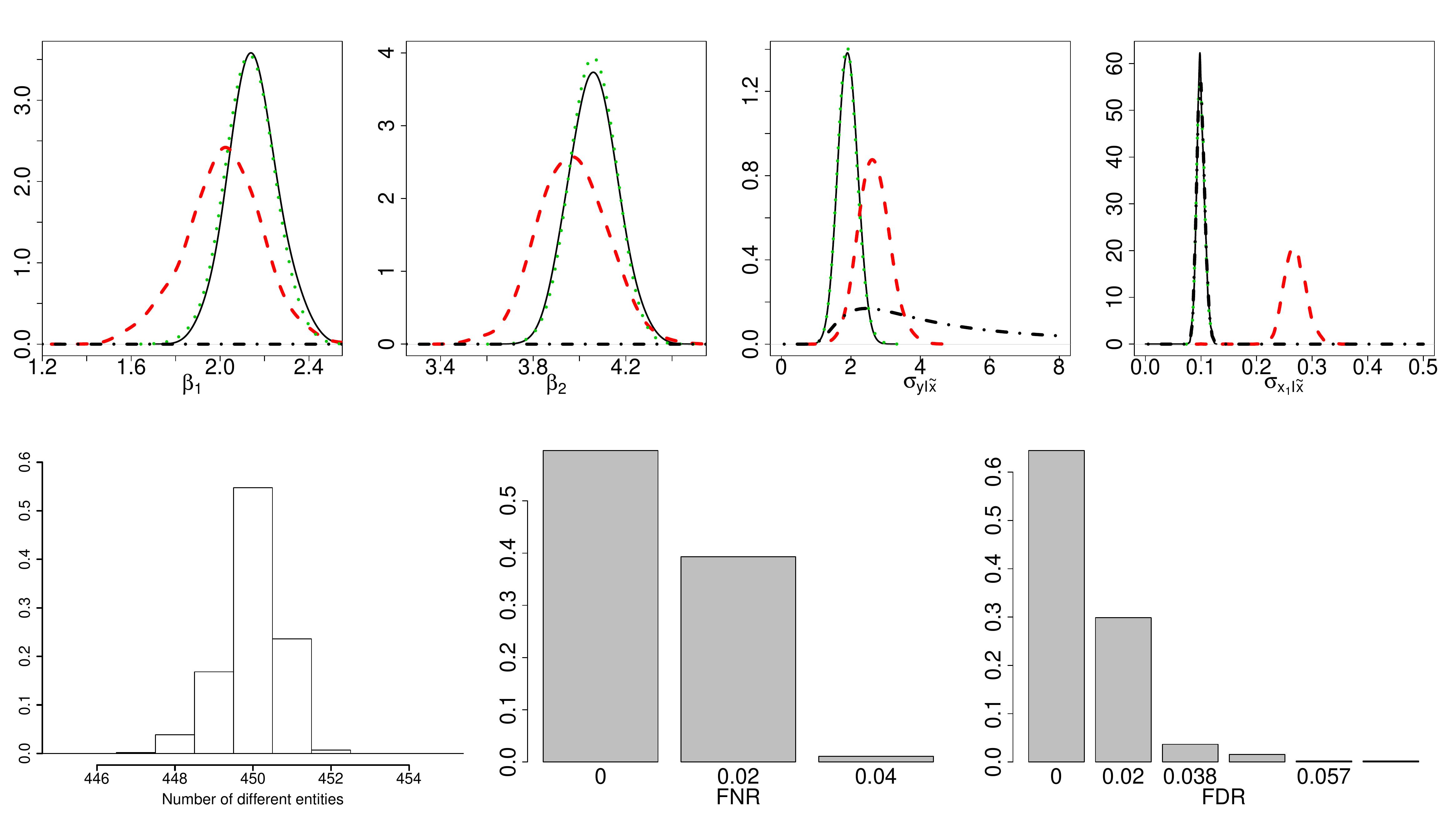}}
\caption{\label{twocov}
Experiment III. Same caption as Figure \ref{complete}.
}
\end{figure}

Figure \ref{complete} gives the results of Experiment I.  The posteriors of $(\beta$, $\sigma_{y|\tilde{x}}$, $\sigma_{x|\tilde{x}})$ from our joint modeling approach (first row, solid lines) do not show remarkable differences  when compared  to their true counterpart (first row, dotted lines), which were obtained by fitting the regression model conditional on the true value of $\lambda.$ 
The similarity between the  posteriors is mainly due to the large concentration of $\lambda$ around the true pattern of duplications. The mode of the posterior of the number of distinct entities is exactly the true value (450), where the FNR and FDR are considerably smaller with respect to case without the $y$ and $x$ columns. Hence, the effect of considering the information provided by the regression model has improved the record linkage process.

Figure \ref{broken} gives the results of Experiment II.
The posteriors (first row, solid lines) of $(\beta, 
\sigma_{y|\tilde{x}}, \sigma_{x|\tilde{x}})$ are similar to the
corresponding true posteriors (first row, dashed lines). 
We report the posteriors obtained by fixing $\lambda$ equal
to the point estimate provided by the hit-and-miss model applied to the
categorical variables alone (first row, dotted lines). 
The posteriors of $\beta$ and
$\sigma^2_{y|\tilde{x}}$ obtained with the plug-in approach are strongly biased for the presence of false matches which, on the other hand, are not affecting the posterior of $\sigma_{x|\tilde{x}}$. This distribution depends
on the 13 duplicated entities with two copies of $x$ which are
correctly accounted for in the plug-in approach. To better illustrate the causes of the distortion in the estimation of the regression parameters, the right panel on the top row shows all the $(x,y)$ pairs resulting from the plug-in approach.  The solid black
circles represent the true matches, and the empty red circles represent
the false matches, with independent $y$ and $x$ values. We 
report the corresponding regression lines, where the three false matches are lowering the $\beta$ estimate and increasing the $\sigma_{y|\tilde{x}}$ estimate.
Further analysis reveals that  the posterior 
for $k(z)$ (second row) with the integrated hit-miss and regression model is less
concentrated with respect to the first experiment but it is more
concentrated with respect to the single hit-miss model. We
reduce the FDR, leaving the FNR almost unchanged.
We coin this the \emph{feedback effect} of the regression from the downstream task. 
For example, if we consider a false link, the posterior probability of being a match will typically be down-weighted by the low likelihood arising from the regression part of the model. Hence, in addition to centering the estimates of the
regression coefficient $\beta$, the joint regression-hit miss model
improves record linkage performance.


Figure \ref{twocov}  gives the results of the Experiment III.
The joint model gives posteriors similar to the true ones
while the plug-in approach gives biased estimates and larger
variability (first row, left upper panels).  
The presence of false matches in the plug-in approach gives a positive bias in estimating the variance $\sigma_{y|\tilde{x}}$ and affects 
the posterior of the measurement error parameters (first row, right upper panels).  The posteriors of $\sigma_{x_1|\tilde{x}}$ and $\sigma_{x_2|\tilde{x}}$ (not reported) both with the joint model  and the true $\lambda$ are essentially equal to the prior, while the plug-in posterior is concentrated on larger values. Under such conditions, even with the true linkage structure,  we do not have  any useful information for estimating the measurement error variances due to the lack of duplicated $x$ values.  Thus, while the joint model correctly does not contrast the information provided by the prior, the presence of false matches creates $(y,x)$ pairs that could be also explained by a larger measurement error of the covariates.  We observe that the joint modeling of the record linkage and regression data improves the matching process as noted by the higher concentration of $k(z)$ (second row, left lower panel) around the true value  of 450 and the lower FNRs and FDRs (second row, right lower panels) with respect to results obtained with the hit-and-miss model only.

\section{Discussion}
\label{sec:discussion}
We have made three major contributions in this paper. First, we have proposed a Bayesian record linkage model investigating the role that  prior partition models may have on the matching process. Second, we have proposed a generalized framework for record linkage and regression that accounts for the record linkage error exactly. Using our methodology, one is able to generate a feedback mechanism of the information provided by the working statistical model on the record linkage process. This feedback mechanism is essential to eliminate potential biases that can jeopardize the resulting post-linkage inference. Third, we illustrate our record linkage and multiple regression methodology on many experiments involving a synthetic data set, where improvements are gained in terms of standard record linkage evaluation metrics.  

\section*{Acknowledgments}
Steorts was supported by NSF-1652431 and NSF-1534412. Tancredi and Liseo were supported by Ministero dell' Istruzione dell' Universita e della Ricerca, Italia PRIN 2015.

\bibliographystyle{ims}
\bibliography{regression}

\section*{Appendix}
\appendix
\normalsize

\section{Metropolis Algorithm}
\label{app:metropolis}
We provide our Metropolis-within-Gibbs algorithm that allows direct simulation from the joint posterior. Let $\lambda_{(-ij)}$ be the vector $\lambda$ with the element $\lambda_{ij}$ removed and let
$C_{j'} \setminus (ij) $ be the set of all the observed records with record $(i j)$ removed, which conditionally on $\lambda$, refer to the population individual $j'$
The full conditional distribution of $\lambda_{ij}$ is
\begin{align}
\label{condlambda}
p(\lambda_{ij}=q\vert \lambda_{(-ij)},\alpha, v)
& \propto
\prod_{j'=1}^{N_{pop}} p(V_{C_{j'}}=v_{C_{j'}}\vert \alpha,\lambda) \, p(\lambda_{ij}=q \vert \lambda_{(-ij)})\\
& \propto \prod_{j'=1}^{N_{pop}} \frac{ p(V_{C_{j'}}=v_{C_{j'}}|\alpha,\lambda)}{p(V_{C_{j'\setminus (ij)}}=v_{C_{j'} \setminus (ij) } \vert \alpha,\lambda)}  p(\lambda_{ij}=q \vert \lambda_{(-ij)}) \notag \\
& \propto
\frac{ p(V_{C_{q}}=v_{C_{q}}\vert \alpha,\lambda)}{p(V_{C_{q}\setminus (ij)}=v_{C_{q} \setminus (ij) } \vert \alpha,\lambda)} p(\lambda_{ij}=q \vert \lambda_{(-ij)}), \end{align}
where  $q=1,\ldots N_{pop}.$
In equation~\ref{condlambda}, set $\lambda_{ij}=q$, which implies that $C_{j'}=C_{j'\setminus ij}$ $\forall j'\neq q.$ It follows  that
$\dfrac{ p(v_{C_{j'}}|\alpha,\lambda)}{p(v_{C_{j'} \setminus ij } |\alpha,\lambda)}=1
\quad \forall j'\neq q.$
When the population entity $q$ represents an already existing cluster given $\lambda_{-(ij)}$, the above ratio can also be written as
$$
\frac{ p(V_{C_{q}}=v_{C_{q}}\vert \alpha,\lambda)}{p(V_{C_{q}\setminus (ij)}=v_{C_{q} \setminus (ij) } \vert \alpha,\lambda)}
= \prod_{l=1}^p \left[\alpha_l \theta_{l\, v_{ijl}} +(1-\alpha_l) \frac{
\prod_{(i_h,j_h)\in C_q \setminus (ij)} \left( (1-\alpha_l) \delta_{v_{i_h j_h l},v_{i j l}} +\alpha_l \theta_{l v_{i_h j_h l}} \right)
}{p(V_{C_{q}\setminus (ij) \, l}=v_{C_{q} \setminus (ij) \,l } \vert \alpha,\lambda)} \right].
$$
When the label $q$ identifies a new cluster, the following simplification is possible:
$$\frac{ p(V_{C_{q}}=v_{C_{q}}\vert \alpha,\lambda)}{p(V_{C_{q}\setminus (ij)}=v_{C_{q} \setminus (ij) } \vert \alpha,\lambda)}
= \prod_{l=1}^p \theta_{l,v_{ijl}}.$$
Note that the posterior $p(\lambda,\alpha|v)$ is invariant with respect to the cluster labels
and that we are only interested in the cluster composition. 
Thus, we can avoid simulating the entire population label distribution, and instead set 
 $q\in \{1,\ldots, N\} $ (since there can be at most $N$ clusters)  and update $\lambda_{ij}$ with the following:
\begin{equation}
q(\lambda_{ij}=q)=\left\{
\begin{array}{c l}
 \frac{ p(V_{C_{q}}=v_{C_{q}}\vert \alpha,\lambda)}{p(V_{C_{q}\setminus (ij)}=v_{C_{q} \setminus (ij) } \vert \alpha,\lambda)}
p(\lambda_{ij}=q|\lambda_{(-ij)}) &
\textnormal{if $q$ labels an observed cluster}
\\
\prod_{j=1}^p \theta_{l, v_{ijl}} p(\lambda_{ij}=new|\lambda_{(-ij)}) /(N-k_{(-ij)}) &
\textnormal{if $q$ labels a new cluster}
\end{array}
\right.
\label{gibbs}
\end{equation}
 for $i=1,\ldots,L, j=1,\ldots, N_i$, where $k_{(-ij)}$ is the number of clusters without the label $\lambda_{ij}$. This way of updating the cluster assignment is standard when the CRP is used for a prior on the cluster assignments. In addition, the marginal likelihood of the cluster observations is known or can be easily calculated using a recursive formula, see for example \cite{maceachern1994} and \cite{neal2000}.
 
To adapt the algorithm to the two different prior distribution of $\lambda$, note that, when $q$ labels an observed cluster, the use of a uniform prior on $\lambda$ implies that
$$p(\lambda_{ij}=q|\lambda_{-(ij)})=1/N_{pop} \quad \textnormal{and} \quad p(\lambda_{ij}=new|\lambda_{-(ij)})=(N_{pop}-k_{-(ij)})/N_{pop}.$$
With the PYP prior, the above mentioned probabilities are, respectively,
$$p(\lambda_{ij}=q|\lambda_{-(ij)})=(n_q-\sigma)/(N-1+\vartheta) \quad p(\lambda_{ij}=new|\lambda_{-(ij)})=(k_{-(ij)}\sigma+\vartheta)/(N-1+\vartheta)$$ where $n_q$ here denotes the size of the cluster $C_q$ without the entity $\lambda_{ij}$.
Finally, when a uniform prior on the partition space is considered, one has
$$p(\lambda_{ij}=q|\lambda_{-(ij)})\propto1/(N_{pop})_{k(-ij)} \quad \textnormal{and} \quad p(\lambda_{ij}=new|\lambda_{-(ij)})\propto(N_{pop}-k_{-(ij)})/(N_{pop})_{k_{(-ij)+1}}.$$

Finally, full conditional distributions of the components of $\alpha$ have a computationally manageable form using a recursive formula. 
In fact, assuming a standard Beta prior on each $\alpha_l$, one obtains
$$p(\alpha_l|\lambda,v,\alpha_{-l}) \propto
\prod_{j'=1}^N p(V_{C_{j'} l} = v_{C_{j'} l }\vert \alpha_l)\alpha_l^{p-1} (1-\alpha_l)^{q-1}, $$
and a straightforward Metropolis step can be easily implemented.

\section{Construction of PYP Priors}
\label{app:pyp}

We now briefly describe adapting the PYP prior to our $L$ database framework. Assume the first  $j$ records of the  $i$-th database and all the records of the first $i-1$ databases   are classified into $k_{i,j}$ clusters   identified  by the the population labels 
$j'_1,\ldots, j'_{k_{i,j}}$ with sizes
$n_1, n_2, \dots, n_{k_{i,j}}$ respectively. Also, let $N_{i,j}=\sum_{l=1}^{i-1}N_l+j$ denote the total number of these records.  Next, the label of the record  $\lambda_{i,j+1}$ identifies a new cluster with probability
$$
\proba{\lambda_{i,j+1}= \mbox{``new''}\vert \lambda_{1, 1}, \dots ,\lambda_{i, j}} = \frac{k_{i,j} \sigma+\vartheta}{N_{i,j}+\vartheta},
$$
where $(\vartheta,\sigma)$ are two parameters whose admissible values are $\sigma \in [0,1)$ with $\vartheta>-\sigma$ or $\sigma<0$ with $\theta= m|\sigma|$ for some positive integer $m$. Moreover, $\lambda_{i,j+1}$ will assume an already observed  label $j'_g$ identifying a cluster  with size $n_g$   with probability
$$
\proba{\lambda_{i,j+1}=j'_g\vert \lambda_{1,1}, \dots ,\lambda_{i_1,j_1} }= \frac{n_g - \sigma }{N_{i,j}+\vartheta} \quad g=1, \dots, k_{i,j}.
$$
The above updating rule induces a prior on the set of the possible partitions of all the  $N$ records which can be written as \citep{pitman}
$$
\proba{z(\lambda)=z } = \frac{(\vartheta + \sigma)_{k-1, \sigma}}
{(\vartheta + 1)_{N-1, 1}} \prod_{g=1}^k (1-\sigma)_{n_g-1, 1},
$$
where $\{n_1, \dots, n_k\}$ are the cluster sizes of the partition $z$ and
$x_{r,s}= x (x+s) \cdots (x+(r-1) s).$
It can also be proved \citep{pitman} that, under this prior, the expected value of $k(z)$ is
$$
E(k(z)) = \sum_{i=1}^N \frac{(\vartheta + \sigma)_{(i-1)\uparrow}}
{(\vartheta + 1)_{(i-1)\uparrow}}
=\frac{\vartheta}{\sigma} \left[ \frac{(\vartheta+\sigma)_{N\uparrow}}{\vartheta_{N\uparrow}}-1\right]$$
and the variance is
$$Var(k(z))=\frac{\vartheta(\vartheta+\sigma)}{\sigma^2}  \frac{(\vartheta+2\sigma)_{N\uparrow}}{\vartheta_{N\uparrow}}
-\frac{\vartheta^2}{\sigma^2}\left( \frac{(\vartheta+\sigma)_{N\uparrow}}{\vartheta_{N\uparrow}} \right)^2-\frac{\vartheta}{\sigma} \frac{(\vartheta+\sigma)_{N\uparrow}}{\vartheta_{N\uparrow}}$$
with $x_{s \uparrow} = \Gamma(x+s)/\Gamma(x)$. For more details, we refer to \cite{yamato2000moments}.

The above equations can be used for prior elicitation by fixing $\vartheta$ and $\sigma$
in order to have $E(k(z))$ equal to a rough prior guess for the number of clusters and a specific amount of prior  variability for $k(z)$. 
Moreover, in evaluating the asymptotic properties, \cite{pitman}   observes  that  as $N \to\infty$, $E(k(z))$ becomes infinite for non negative values of $\sigma$; on the other hand, if $\sigma$ is negative,  $k(z)$ is equal almost surely to $m$ which thus takes the role  of the size $N_{pop}$
in a finite population framework.  

\section{Record linkage without duplicate-detection}
\label{app:duplicate-detection}
We now consider record linkage of two databases without duplicate-detection. 
To consider this case, we simply modify the
prior distribution on the $\lambda$'s such that
$\lambda_{i j_1}\neq \lambda_{i j_2}$ $\forall j_1\neq j_2$ and for $i=1,2$.
In this case, clusters consist of at most two elements so that the distribution of the observed records $v$, conditional on $\lambda$ and $\alpha$, can be calculated analytically without exploiting the recursive formula.

If a uniform prior on the label space is assumed, the above conditioning is equivalent to assuming that the two databases are two simple random samples with replacement from a population of $N_{pop}$ units. This is  the same situation described in \cite{tancredi:liseo2011}, where $N_{pop}$ is assumed unknown. Assume that $T$ denotes the number of common units between the two databases; then  $k(z)$ is equal to  $N_1+N_2-T$, where $T$ follows a hypergeometric distribution
$$P(T=t)=\frac{\binom{N_1}{t} \binom{N_{pop}-N_1}{N_2-t}}{\binom{N_{pop}}{N_2}}, \quad   \max\{0, N_1+N_2-N_{pop} \}\leq  t\leq \min\{N_1,N_2\}.$$

From a computational perspective, the conditioning of the uniform prior does not imply substantial changes. In fact
if a PYP prior is assumed, the standard record linkage framework can be tackled by imposing that
$\lambda_{1j}=j$ for $j=1,\ldots, N_1$ and that the units of the second database may only join a cluster composed by a single unit of the first database or create a new cluster, that is
$$p(\lambda_{2\, j+1}=q|\lambda_{11}\ldots, \lambda_{2\,j})=
\left\{
\begin{array}{c c}
\frac{1-\sigma}{k_{2 j}-j(1-\sigma+\vartheta)} & \textnormal{ if } q\leq N_1 \textnormal{ and } n_q=1\\
0 & \textnormal{ if } q\leq N_1 \textnormal{ and } n_q=2\\
0 & \textnormal{ if } q>N_1 \textnormal{ and } n_q=1, \\
\end{array}
\right. \quad j=0,1\ldots, N_2-1
$$
 and
$$p(\lambda_{2\, j+1}=new|\lambda_{11}\ldots, \lambda_{2\,j})=
\frac{k_{2j} \sigma+\vartheta}{k_{2j}-j(1-\sigma)+\vartheta} \quad j=0,1\ldots, N_2-1 $$
where $k_{20}=N_1$ and $k_{2j}$ is the number of distinct elements considering the first database and the first $j$ elements of the second database.
Finally, notice that
$$p(\lambda_{2 1},\ldots, \lambda_{2 N_2}|\lambda_{11},\ldots,\lambda_{1 N_1})=\frac{(1-\sigma)^{N-k_{2 N_2}} \prod_{l=N_1+1}^{k_{2 N_2}} (\sigma (l-1)+\vartheta ) } {\prod_{l=1}^{N_2} (k_{2\, l-1}-(l-1)(1-\sigma)+\vartheta ) } \frac{(N-k_{2 N_2})!}{N!}.$$
This implies that the $\lambda$'s are no longer exchangeable. This problem, although interesting from a theoretical perspective, does not cause computational issues.

The conditional prior probabilities for the Gibbs step updating of $\lambda_{2j}$ to be used from equation (\ref{gibbs}) are
$$p(\lambda_{2j}=q |\lambda_{-(2j)}) \propto
\left\{
\begin{array}{c c}
(1-\sigma) & \textnormal{ if } q\leq N_1 \textnormal{ and } n_q=1\\
0 & \textnormal{ if } q\leq N_1 \textnormal{ and } n_q=2\\
0 & \textnormal{ if } q>N_1 \textnormal{ and } n_q=1, \\
\end{array}
\right.
$$
and

$$p(\lambda_{2j}= new | \lambda_{-(2j)}) \propto (k_{-(2j)}\sigma+\vartheta) \prod_{l=j}^{N_2-1} \left[
\frac{k_{2l}-l (1-\sigma) +\vartheta}{k_{2l}+1-l(1-\sigma) +\vartheta}
\right].
$$

\section{Record Linkage Experiment}
\label{app:record-linkage}
We provide the parameter settings for the record linkage experiments. 
For the case with duplicate detection, we considered the effect of the PYP prior for $\lambda$ with $(\theta,\sigma)$=(0.4,0.98), (2,0.975), (10,0.965). These prior distributions  have a common prior mean of $k(z)$ almost equal to 450; however, their respective variance are quite different. For the case of no duplicate detection, 
 we consider the effect of the constrained  PYP prior for $\lambda$  with $(\theta,\sigma)= (1,0.6),$ $(1,0.725),$ and  $(1,0.86).$   These values of the  hyper-parameters $(\theta,\sigma)$ produce prior means for the number of matches equal to 75, 50 and 25.

\section{Regression Experiments}
\label{app:regression}

We elaborate on our three regression experiments. In the first experiment, we modify the data set by adding two columns with the pairs $y$ and $x$ generated from the model in section \ref{sec:regression-simp}, conditional on the true $\lambda$ structure. For clusters with two records we simulate a single true value $\tilde{x}$ of the covariate from a normal distribution with zero mean and variance equal to $\sigma^2_{\tilde{x}}=9.$  Then, conditionally on $\tilde{x},$ we generate two independent draws $x$ from a normal distribution with mean $\tilde{x}$ and variance $\sigma^2_{x|\tilde{x}}=0.01$ and  two corresponding independent draws $y$ from a normal distribution with mean $\beta \tilde{x}$ with $\beta=3$ and variance $\sigma^2_{y|\tilde{x}}=4$. Instead, the records without duplication are augmented with a single pair $(y,x)$ that is generated conditionally on a single value $\tilde{x}$ following the same model of the duplicated records. 

In the second experiment, we use the modified \texttt{RL500} data set that consists of two databases. We then remove $y$ from the
second databases and $x$ from the first database. Given the 50 entities with duplication, 28 
belong to both the databases reporting the $y$ variable on the first database
and the $x$ variable on the second database. Moreover, 9 entities only belong to first database
with 2 duplicate records of $y,$  and 13 entities only belong to the second database
with 2 copies of $x.$ In addition, we assume the same priors as in the first experiment.




In the third experiment, we modify the \texttt{RL500} data set by generating data from a regression model with two covariates, where we assume $\beta_1=2$ and $\beta_2=4$, $\sigma^2_{y|\tilde{x}}=4$ and a diagonal covariance matrix
$\Sigma_{x|\tilde{x}}$ with elements  $\sigma^2_{x_1|\tilde{x}}=\sigma^2_{x_2|\tilde{x}}=0.01.$ We then split this data set into two databases of size 250, and then remove $y$ from the second database and remove the two covariates from the first database. To mimic  the case of record linkage without duplicate detection we arrange the two databases so that they  share 50 entities without duplications within each databases.


\end{document}